\documentclass[useAMS,usenatbib]{mn2e}

\def\apj{ApJ}                 
\def\aap{A\&A}                
\def\aapr{A\&A~Rev.}          
\def\mnras{MNRAS}             

\usepackage{epsf}

\title[Self-Similar Magnetic Arcades]{Self-Similar Magnetic Arcades}
\author[K.N. Gourgouliatos]{K.N. Gourgouliatos\thanks{E-mail:
kng22@ast.cam.ac.uk} \\
Institute of Astronomy, University of Cambridge, Madingley Road CB3 0HA}
\begin{document}

\date{Accepted 2007 December 13. Received 2007 December 5; in original form 2007 October 23}

\pagerange{\pageref{firstpage}--\pageref{lastpage}} \pubyear{-}

\maketitle


\label{firstpage}
\begin{abstract}
We study self-similar analytical solutions for force-free magnetic field in azimuthal symmetry and arcade topology. We assume the existence of a poloidal magnetic field, anchored on a heavy spherical conductor. The field is changed by shearing the foot points of the arcade due to differential rotation. This rotation gives rise to a toroidal component in the magnetic structure which reacts by expanding the poloidal flux outwards. This could be a slow process at the early stages, however it becomes very fast at the final stages when the poloidal flux expands to infinity. We address the question of the pressure environment confining the arcade, a pressure profile proportional to $r^{-4}$ is particularly interesting as it allows finite twist before the field expands to infinity. Finally, some time evolution estimates are made to demonstrate the limitations of this study.
\end{abstract}


\section{Introduction}

Previous work by \cite{1992ApJ...391..353W, 1994MNRAS...267..146L, 1994A&A...288.1012A} studied the effect of twist on a magnetic structure. \cite{1992ApJ...391..353W} studied the case of a force-free magnetic dipole that is sheared at the stellar surface, expands and opens partially. \cite{1994MNRAS...267..146L} studied the case of a quadruple force-free magnetic field anchored on a heavy conductor that expands outwards because of differential rotation at the base and found that after finite twist the field opens totally and takes the topology of a split monopole. \cite{1994A&A...288.1012A} studied a 2-D model of arcade topology where the magnetic field expands because of the motion of the foot points at the surface where the magnetic field emerges.  

In this paper we study the problem of force free magnetic arcades of azimuthal symmetry in three dimensions using the methods developed by \cite{1994MNRAS...267..146L} and we seek for analytical solutions confined between two cones in order to form an arcade. It is not very likely that we can find solutions that will provide essential physical insight by just a numerical solution of the partial differential equation. It proves more interesting to design a problem that leads to analytical solutions. The method we are going to use can be described as a ``backwards method'', namely we are going to assume axial symmetry and self-similarity. Then we are going to evaluate how much differential rotation we must impose at the surface in order to achieve this structure and what is the external pressure environment that balances the magnetic pressure. This method is somewhat unnatural because it inverts the cause and the result, nevertheless it simplifies the problem and gives analytical solutions to a highly non-linear MHD problem. The alternative ``forward method'' of attacking the problem is more physical, it gives a wider class of solutions but the accuracy of its results is limited due to the forms of trial functions employed in the solution. In the ``forward method'' we take a given pressure environment, some flux emerging from the base and some differential rotation and we have to determine the shape of the magnetic cavity and the details of the magnetic field inside the cavity.

\cite{1991ApJ...375L..61A} has pointed out that there is a maximum energy that can be stored in a field whose foot points are fixed on a sphere and an upper bound is given by the field that goes straight outward to infinity. However in the case of a plane arcade in two dimensions no such bound exists and indeed such a two dimensional arcade can be sheared by any amount so that the field energy can be arbitrarily large. One of our aims is to demonstrate that arcades around the sun start as in the Aly 2-D case but once they have expanded enough to encounter the significantly larger areas available at great radii they rapidly increase in height and the flux expands to infinity after only a finite shear, unlike the 2-D case but in perfect conformity to Aly's bound. This upper energy bound has been proven for the case of a field occupying the space outside a sphere. Following Aly's proof step by step, it can be extended to the arcade case, where the field is confined by a sphere and two cones, provided that there is no normal component of the field into the bounding cones. Indeed the configuration contains a current sheet at the conical boundaries. The pressure is continuous, as the external pressure is equal to the magnetic pressure at the boundary, for the arcade to be in equilibrium.

\section{The Problem}

In this section we state the physical properties of the configuration we are studying. A well defined framework is essential for analytical solutions. 

\subsection{The Grad-Shafranov equation}

Consider a magnetic field of axial symmetry rising from and returning to the surface of a sphere but confined between two cones of semi-opening angles $\theta_{1}$ and $\theta_{2}$ (Fig. 1). The magnetic field is force-free and obeys the equations of ideal MHD. In what follows the symbols have their usual meaning. The force-free condition is

\begin{eqnarray}
 \mathbf{j} \times \mathbf{B}=0.
\end{eqnarray}

\begin{figure}
\begin{center} 
\epsfxsize=8.7cm 
\epsfbox{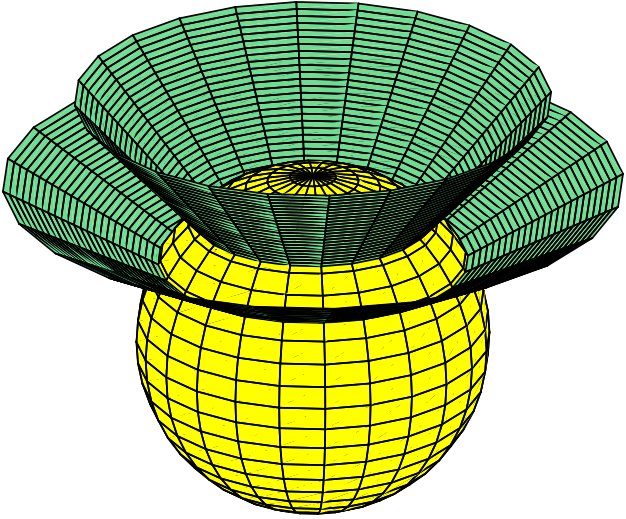}
\caption[]{The geometry of the problem. The field is confined in the space outside the sphere and between the two cones.}
\label{}
\end{center}
\end{figure}
 
The non-relativistic version of the induction equation, that is ignoring the displacement current, is

\begin{eqnarray}
 \nabla \times \mathbf{B}=\frac{4 \pi}{c} \mathbf{j}.
\end{eqnarray}

Therefore, substituting equation (2) to equation (1)

\begin{eqnarray}
 (\nabla \times \mathbf{B}) \times \mathbf{B}=0.
\end{eqnarray}

Equation (3) is equivalent to the condition that the curl of the magnetic field is parallel to the magnetic field,

\begin{eqnarray}
\nabla \times \mathbf{B}= \alpha \mathbf{B}.
\end{eqnarray}

The magnetic field is always divergence free, $\nabla \cdot \mathbf{B}=0$ and we can express it in terms of the flux $P(r,\theta)$, through the cup of a sphere which lies nearer to the axis than $\theta$: 

\begin{eqnarray}
\mathbf{B}=\frac{1}{2 \pi r \sin \theta} \nabla P \times \hat{e}_{\phi}+B_{\phi}\hat{e}_{\phi}. 
\end{eqnarray}

From \cite{1999stma.book.....M} the toroidal component of the field can be expressed in the following way:

\begin{eqnarray}
 B_{\phi}=\frac{\beta(P)}{2 \pi r \sin \theta}.
\end{eqnarray}

We substitute expressions (5) and (6) into equation (4) and rewrite the angular derivatives in terms of $\mu=\cos\theta$. We write equation (4) in component form, therefore $P(r,\mu)$ and $\beta(P)$ must obey the following relations

\begin{eqnarray}
 r^2 \frac{\partial^{2} P}{\partial r^2} + (1-\mu^2)\frac{\partial^{2} P}{\partial \mu^{2}}=-r^2 k \beta(P),
\end{eqnarray}

\begin{eqnarray}
 k=\frac{d\beta(P)}{dP}.
\end{eqnarray}

The combination of (7) and (8) leads to the Grad-Shafranov equation in spherical coordinates under axial symmetry\citep{{1958}, {1966}}

\begin{eqnarray}
 r^2 \frac{\partial^{2} P}{\partial r^2} + (1-\mu^2)\frac{\partial^{2} P}{\partial \mu^{2}}=-r^2 \frac{d\beta(P)}{dP} \beta(P).
\end{eqnarray}

Our objective is to solve equation (9) for $P(r,\mu)$, for a certain form of the $\beta(P)$.

\subsection{Self-similar solutions}

The solution to the above partial differential equation depends on the choice of the form of $\beta(P)$. There is little hope to find an analytical solution to the above equation unless we determine $\beta(P)$. Nevertheless we can look for separable solutions which are self-similar in $r$. We express the flux $P$ as a product function of a power law on $r$ and a function determining the angular part $f(\mu)$. We normalise the maximum of $f(\mu)$ to unity using the total flux $F$; $a$ is the radius of the sphere where the field emerges 

\begin{eqnarray}
 P(r,\mu)=F\Big(\frac{r}{a}\Big)^{-l}f(\mu).
\end{eqnarray}

The demand for self-similar solutions constrains to a power law dependence for the source function $\beta(P)$ on $P$ \citep{1994MNRAS...267..146L}. We define $\beta(P)$ in the following way

\begin{eqnarray}
\beta(P)=c_{1}P|P|^{1/l}. 
\end{eqnarray}

Let $c_{0}=c_{1}F^{1/l} a$; the angular part of $P(r,\mu)$ satisfies

\begin{eqnarray}
 l(l+1)f(\mu)+(1-\mu^{2})f''(\mu)=-c_{0}^{2}(1+\frac{1}{l})f(\mu)|f(\mu)|^{2/l}.
\end{eqnarray}

Therefore the non-linear partial differential equation (9) reduces to an ordinary differential equation, which is still non-linear, subject to boundary conditions and the constraint $f(\mu) \le 1$. In the next sections we are going to investigate solutions with arcade topology, but before that we will discuss the form of the fields that appear in these configurations.

\subsection{The fields}

According to the relations introduced in sections 2.1 and 2.2, it is sufficient to determine the form of $f(\mu)$, by solving equation (12). We then obtain detailed expressions for the components of the magnetic fields and consequently the structure of the field lines. The poloidal component of the field is the first term of the right hand side of expression (5), the toroidal is given by expression (6), and by virtue of (10) and (11) the fields become: 

\begin{eqnarray}
 B_{r}=\frac{F a^{l} r^{-l-2}}{2 \pi}\frac{df(\mu)}{d\mu},
\end{eqnarray}

\begin{eqnarray}
 B_{\theta}=\frac{F a^{l} l r^{-l-2}}{2 \pi (1-\mu^{2})^{1/2}} f(\mu),
\end{eqnarray}

\begin{eqnarray}
 B_{\phi}=\frac{c_{0}F a^{l}r^{-l-2}}{2 \pi (1-\mu^2)^{1/2}}f(\mu)|f(\mu)|^{1/l},
\end{eqnarray}

The field lines are determined by the relation:

\begin{eqnarray}
 \frac{dr}{B_{r}}=\frac{r d \theta}{B_{\theta}}=\frac{r \sin \theta d \phi}{B_{\phi}}.
\end{eqnarray}

In order to evaluate the twist of a field line, we are using the equation involving the $\theta$ and $\phi$ components of the magnetic field

\begin{eqnarray}
\frac{r d \theta}{l f}=\frac{r \sin \theta d \phi}{c_{0}f|f|^{1/l}}.
\end{eqnarray}
 
Thus we can integrate and evaluate the total twist on a field line that emerges from a point at $r=a$, $\mu=\mu_{1}$ and ends at $r=a$ and $\mu=\mu_{2}$.

\begin{eqnarray}
\Phi=-\int_{\mu_{1}}^{\mu_{2}}\frac{c_{0} |f|^{1/l}}{l}\frac{d \mu}{1- \mu^{2}}.
\end{eqnarray}

Thus, if we know the form of $f(\mu)$ we have all the information about the field lines.

\section{Arcade Solutions}

In this section we study the magnetic arcades in two distinct cases. The first one is a single arcade bounded between two cones. We increase the twist of the field lines at the base and consequently the arcade expands radially. We evaluate the twist of the field lines and the energy that is stored in this configuration. The second one is a triple arcade. The boundary condition confines  the edges of the external structure, while the central arcade, which is initially identical to the single arcade we study in the first paradigm, has limited mobility. We are interested in this configuration in order to mitigate the effect of the artificial conical boundaries and have a comparison with the single arcade. 

\subsection{The single arcade}

We solve equation (12) to construct a model for a single arcade. The field emerges from a spherical surface of radius $a$ and is constrained between two co-axial cones, whose vertices coincide with the centre of the sphere. Thus the boundary conditions of this configuration, applicable to equation (12) are $f(\mu_{1})=f(\mu_{2})=0$, where $\mu_{1}$ and $\mu_{2}$ are the cosines of the semi-opening angles of the cones. The angular function $f(\mu)$ is normalised to unity, thus we constrain the value of $c_{0}$ for given $l$. We start from $l=9$ and $c_{0}=0$, then we solve the differential equation for $l$ decreasing, so that the structure expands (Fig. 2). We evaluate the total twist of the structure, and we can estimate the differential rotation we need to impose at each field line in order to achieve this configuration (Fig. 3). 

For $c_{0}=0$ the field is purely poloidal and the solution takes the following analytical form

\begin{eqnarray}
 f_{9}=\frac{1}{2.188}(1-\mu^{2})P_{9}'(\mu).
\end{eqnarray}

Where $P'_{9}(\mu)$ is the derivative with respect to $\mu$ of the ninth Legendre polynomial.

\begin{figure}
\begin{center} 
\epsfxsize=8.7cm 
\epsfbox{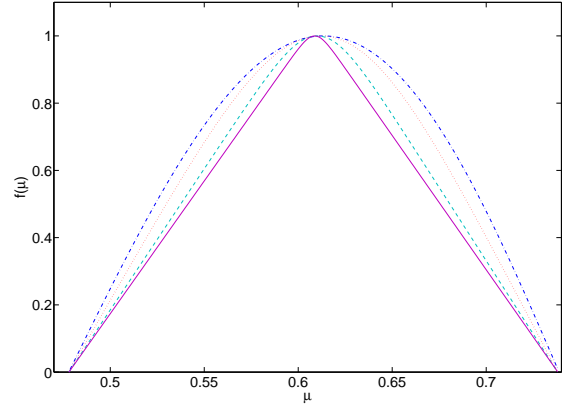}
\caption[]{The angular function $f(\mu)$ of the single arcade for $l$ 9, 1, 0.2, 0.005, dashed-dotted, dotted, dashed and solid lines respectively.}
\label{}
\end{center}
\end{figure}

\begin{figure}
\begin{center} 
\epsfxsize=8.7cm 
\epsfbox{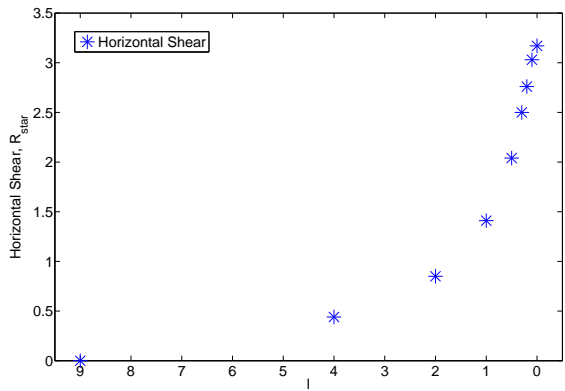}
\caption[]{The dependence of the horizontal shear on the power law index $l$. Only finite differential rotation is needed for the arcade to reach infinity with $l=0$.}
\label{}
\end{center}
\end{figure}

The choice of $l=9$ in the initial case yields that the boundaries of the arcade are $\mu_{1}=0.478$ and $\mu_{2}=0.738$. The other cases are solved numerically and the results appear at table 1. When $l\to 0$ we can use the approximate method developed by \cite{1994MNRAS...267..146L} in order to obtain an analytical expression for $f(\mu)$

\begin{eqnarray}
f(\mu)=1-\frac{l}{1+l}\ln \cosh \Big(\frac{(l+1)c_{0}}{l(1-\mu_{max0}^{2})^{1/2}}(\mu-\mu_{max0})\Big).
\end{eqnarray}

The quantities appearing equation (20) are determined by the asymptotic behaviour of the differential equation (12). As $\cosh$ is an even parity function, in order to satisfy the boundary condition at $\mu_{1}$ and $\mu_{2}$, $\mu_{max}$ obeys $\mu_{1}-\mu_{max0}=-(\mu_{2}-\mu_{max})$ therefore $\mu_{max0}=(\mu_{1}+\mu_{2})/2$. Then it is possible to find $c_{0}$ using again the boundary condition $f(\mu_{1})=0$, thus $c_{0}=2\frac{l\sqrt{1-\mu^{2}_{max0}}}{(l+1)(\mu_{2}-\mu_{1})}cosh^{-1}(e^{(1+1/l)})$, in the limit of $l\to 0$ it becomes $c_{0}=2(1-\mu^{2}_{max0})^{1/2}/(\mu_{2}-\mu_{1})$. The boundary condition $f(\mu_{2})=0$ is satisfied because of the symmetry of the function $f$ with respect to $\mu_{max0}$. The numerical values for $c_{0}$ and $\mu_{max}$ are on tables 1 and 2. A higher order approximation can be done by applying the more sophisticated method developed by \cite{2006MNRAS...369.1167L} (Appendix A), that computes the variation of $\mu_{max}$ as it reaches its asymptotic value. The twist for the limiting case $l \to 0$ can be evaluated analytically as described in the appendix. 

\begin{table}
 \centering
 \begin{displaymath}
\begin{array}{|l|l|l|r|r|l|}
\hline
l & c_{0} & \mu_{max}& f'(\mu_{1})         & \Phi &  Energy \\ 
\hline
9   & 0.00   & 0.613     & 11.35        &  0.00 &  1.00\\ 
4   & 7.87   & 0.613     & 10.87        & -0.71 &  1.97\\ 
2   & 8.22   & 0.613     & 10.35        & -1.30 &  3.22\\ 
1   & 7.87   & 0.612     & 9.72         & -1.99 &  4.68\\ 
0.5 & 7.35   & 0.611     & 9.08         & -2.79 &  6.17\\ 
0.2 & 6.76   & 0.610     & 8.40         & -3.31 &  7.18\\ 
0.1 & 6.47   & 0.609     & 8.08         & -3.61 &  7.65\\ 
0.0 & 6.11   & 0.608     & 7.69         & -3.96 &  8.12\\
\hline
 \end{array}
 \end{displaymath}

 \caption{The results of the numerical solution for equation (12). $\Phi$ is total twist about the axis of the star of the field lines emerging and ending at the boundaries of the arcade corresponding to $P=0$ and the energy is scaled to the energy of the purely poloidal arcade. Expansion to infinity occurs when the twist is somewhat over half a turn.}
 \label{Table 1}
\end{table}

The numerical solution of equation (12) reveals the properties of the magnetic configuration. The increase of the twist amplifies the energy stored in the arcade by two means, the $\phi$ component of the magnetic field becomes stronger and the arcaded expands. As opposed to the plane-parallel model developed by Aly, finite twist leads to disconnection of the arcade into an outgoing and an incoming part. The $B_{\phi}$ component concentrates at the cone separating the incoming and the outgoing part of the arcade and a sheet current forms in the $\hat{e}_{\phi}$ direction. The energy stored in the configuration increases, but is always finite. The foot points of the field lines move weakly in latitude, but somewhat more significantly close to the centre of the arcade.

\subsection{The triple arcade}

In order to mitigate the effect of the external boundaries on the arcade we consider three neighbouring arcades, where the central one has initially exactly the same structure as the arcade of the previous section. At the untwisted stage, that is $l=9$ and $c_{0}=0$; the other two are determined by the Legendre Polynomial we chose. Then, following the same process the foot points are differentially rotated in the azimuthal direction leading to the expansion of the field. The advantage of this method is that the artificial boundaries have been removed from the arcade we are interested in and we have allowed limited mobility. The whole structure is confined again between two cones of cosines of semi-opening angles $\mu_{3}$ and $\mu_{4}$ that determine the boundary condition $f(0.165)=f(0.920)=0$, these correspond to the first and the fourth root of the expression of $(1-\mu^{2})P_{9}'(\mu)$; we reserve $\mu_{1}$ and $\mu_{2}$ for the boundaries of the central arcade (table 1). We have chosen again $l=9$ in the initial case so that the solutions have three humps, and in order to be able to compare directly the triple to the single arcade (Fig. 4).

\begin{figure}
\begin{center} 
\epsfxsize=8.7cm 
\epsfbox{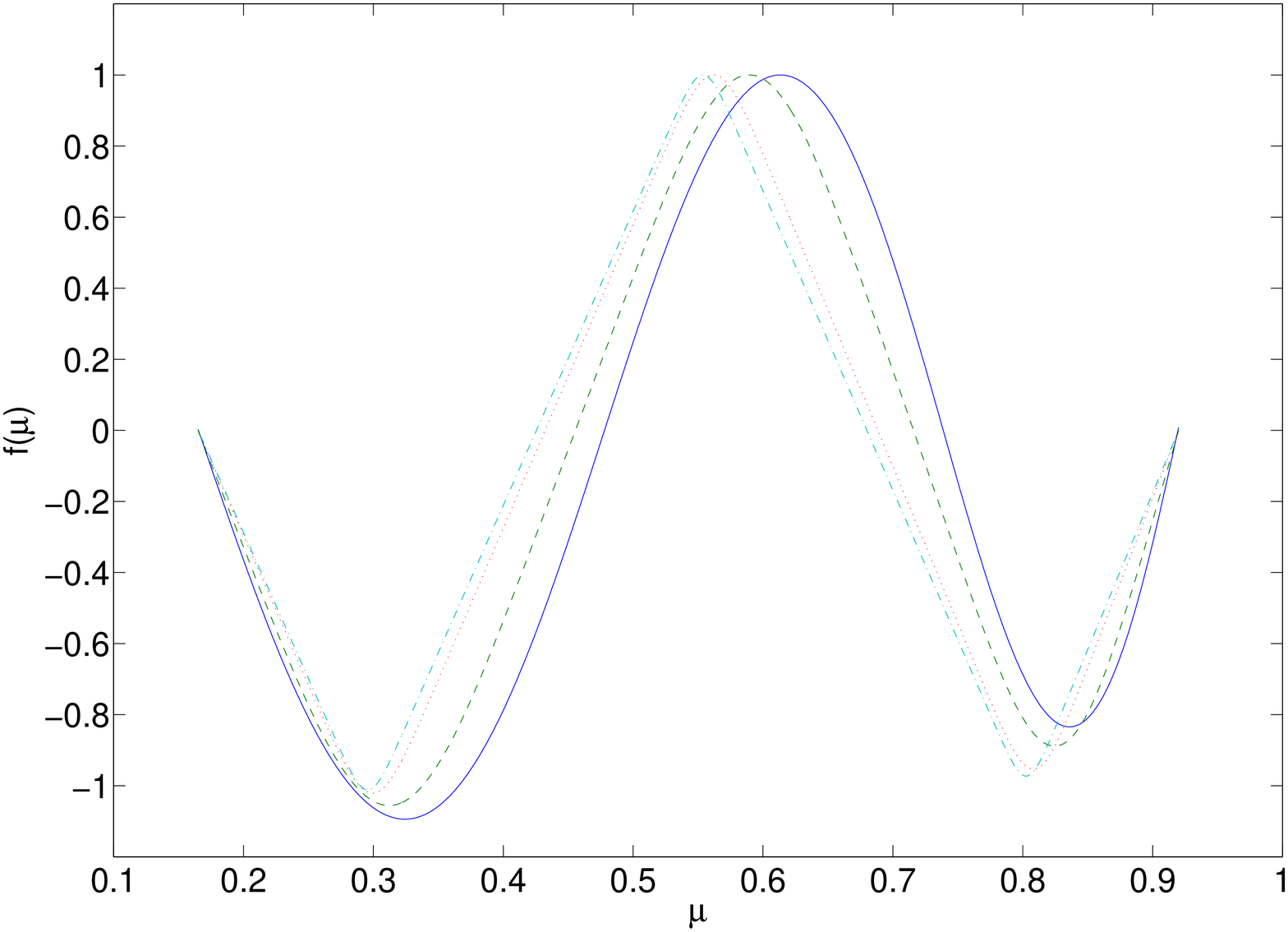}
\caption[]{The angular function $f(\mu)$ of the triple arcade with $l$= 9, 1, 0.2, 0.1, solid, dashed, dotted, dashed-dotted respectively..}
\label{}
\end{center}
\end{figure}

Following the same process we gradually decrease $l$ down to zero and we solve numerically the problem under the normalisation $f_{max}=1$. The results appear in table 2.

\begin{figure}
\begin{center} 
\epsfxsize=8.7cm 
\epsfbox{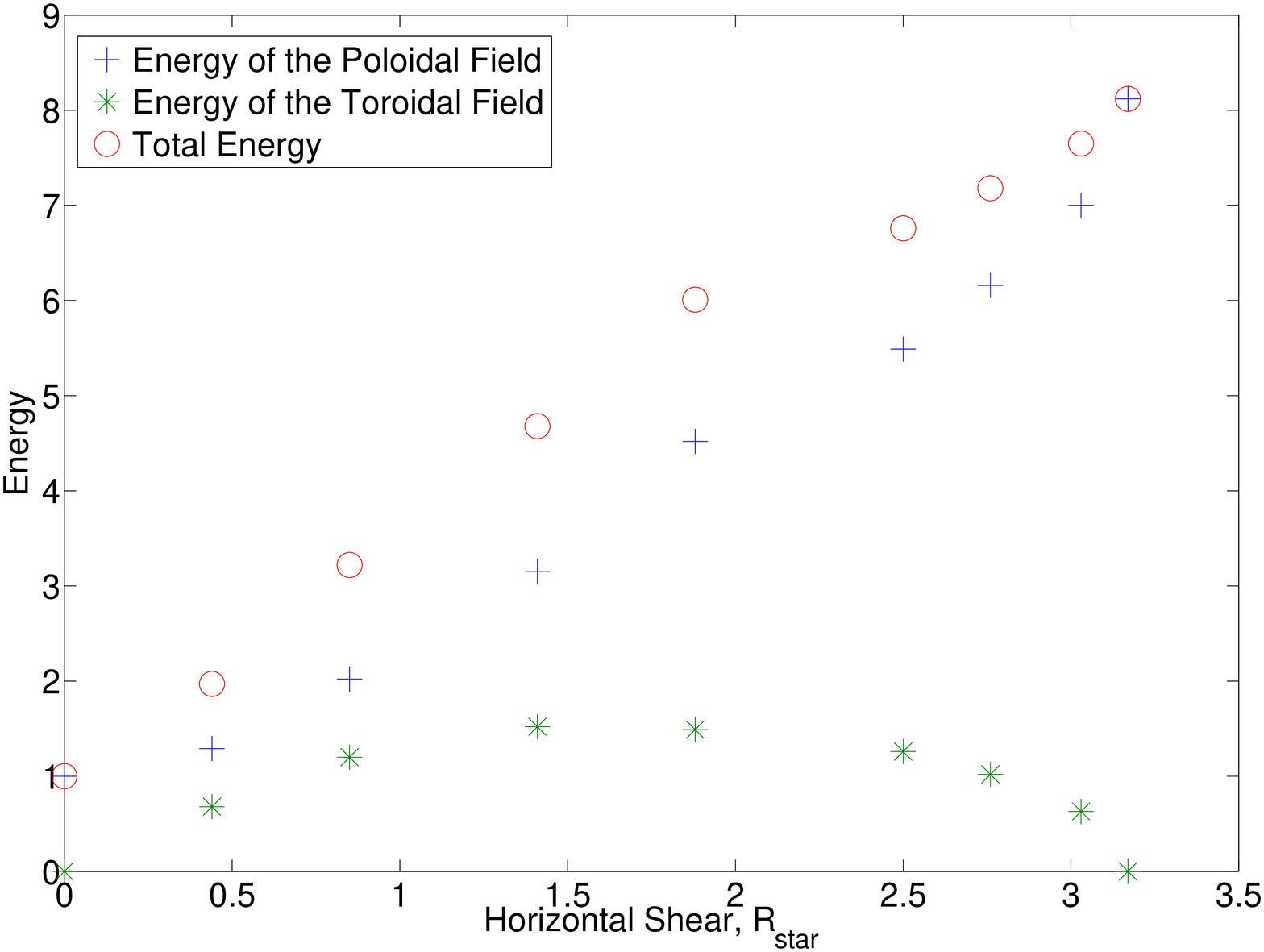}
\caption[]{The energy of the the arcade. Initially, the energy is only due to the poloidal component. As the shear increases there is energy in the toroidal component as well, however, after some point the toroidal component moves to greater distances from the base and its contribution to the total energy decreases. In the final state of the split monopole the energy is only due to the poloidal component, as expected.}
\label{}
\end{center}
\end{figure}

\begin{table}
 \centering
 \begin{displaymath}
\begin{array}{|l|l|l|l|l|l|r|r|}
\hline
l   & c_{0} & \mu_{1}& \mu_{2}& \mu_{max}& f'(\mu_{1}) & \Phi & Energy \\ 
\hline 
9   & 0.00 & 0.478  & 0.738  & 0.613    & 11.35      &  0.00 & 1.00 \\ 
4   & 7.88 & 0.472  & 0.733  & 0.607    & 10.89      & -0.71 & 1.98\\ 
2   & 8.29 & 0.465  & 0.726  & 0.600    & 10.33      & -1.28 & 3.20\\ 
1   & 8.03 & 0.456  & 0.716  & 0.589    & 9.74       & -1.94 & 4.66\\ 
0.5 & 7.81 & 0.445  & 0.704  & 0.577    & 9.15       & -2.64 & 5.88\\ 
0.2 & 7.16 & 0.432  & 0.688  & 0.562    & 8.55       & -3.17 & 7.26\\ 
0.1 & 6.93 & 0.426  & 0.680  & 0.554    & 8.28       & -3.43 & 7.79\\ 
0.0 & 6.69 & 0.417  & 0.668  & 0.543    & 7.94       & -3.74 & 8.36\\
\hline
\end{array}
 \end{displaymath}

 \caption{The numerical solution of the triple arcade. The energy refers to the energy of the central arcade and $\Phi$ to the twist of the field line that lies in the central arcade and has $P=0$. There is minor deviation from the results for the single arcade appearing at table 1 because the foot points now move a little in lattitude.}
 \label{Table 2}
\end{table}

The overall image is similar to the previous case, the arcade expands outwards as we increase the twist of the foot points and energy is stored in the magnetic configuration (Fig. 5). The angular size of the central arcade varies very weakly with $l$, although both foot points move towards the equator and the twist is concentrated in the centre of the arcade (Figures. 6-10). 

\begin{figure}
\begin{center} 
\epsfxsize=8.7cm 
\epsfbox{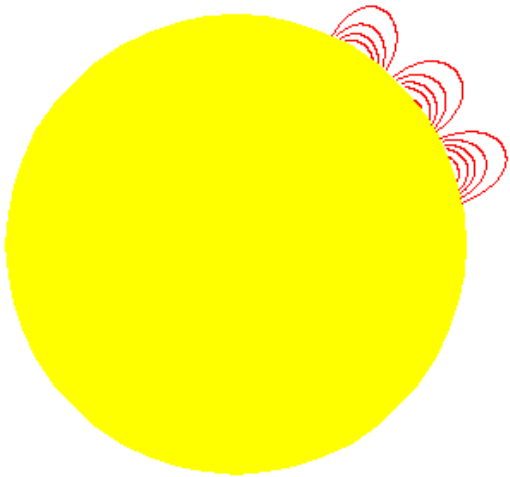}
\caption[]{The magnetic field lines of the triple arcade as seen from the side in the untwisted stage for $l=9$.}
\label{}
\end{center}
\end{figure}

\begin{figure}
\begin{center} 
\epsfxsize=8.7cm 
\epsfbox{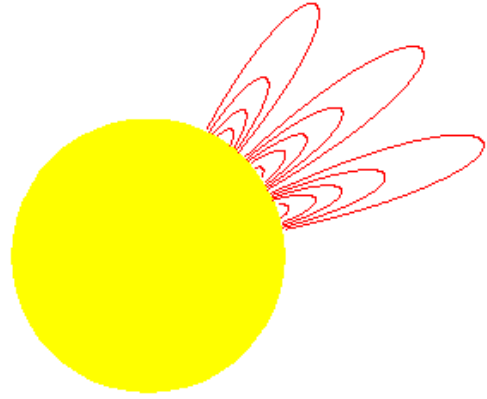}
\caption[]{The magnetic field lines of the triple arcade as seen from the side for $l=2$, the twist has caused a significant expansion.}
\label{}
\end{center}
\end{figure}

\begin{figure}
\begin{center} 
\epsfxsize=8.7cm 
\epsfbox{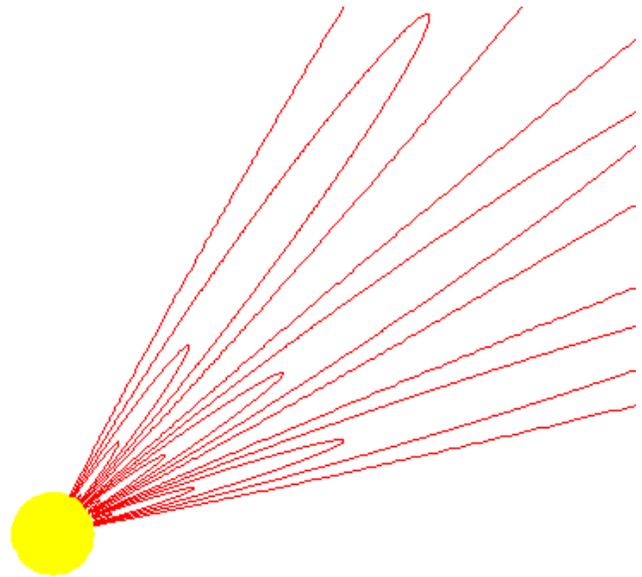}
\caption[]{The magnetic field lines of the triple arcade as seen by the side for $l=0.4$, a little before the field lines open.}
\label{}
\end{center}
\end{figure}

The solutions of the single and the triple arcade give the essential understanding to this problem. We see little deviation between the two models, therefore mobility in the boundaries does not alter significantly the general properties, apart from some minor changes in the total energy and the total twist of the configuration. Therefore in future and more detailed studies we can neglect minor movements in latitude at the base and keep the boundaries fixed.

\begin{figure}
\begin{center} 
\epsfxsize=8.7cm 
\epsfbox{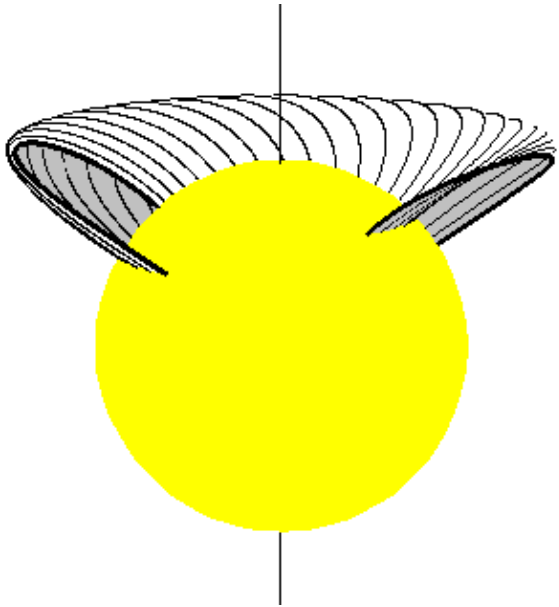}
\caption[]{The twist of the magnetic field lines of a surface of constant flux, for $l=2$.}
\label{}
\end{center}
\end{figure}

\begin{figure}
\begin{center} 
\epsfxsize=8.7cm 
\epsfbox{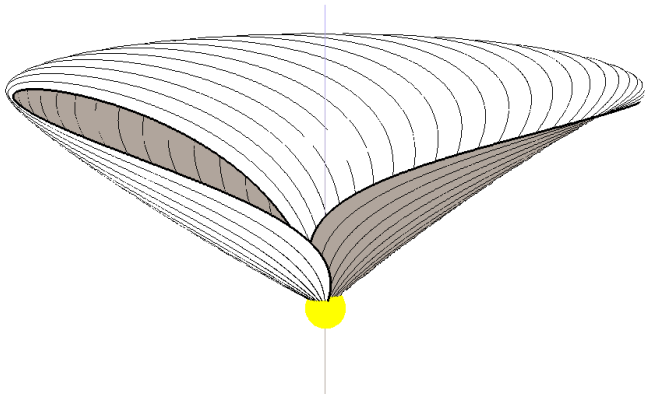}
\caption[]{The twist of the magnetic field lines for $l=0.4$, the expansion of the configuration and the increase in the twist are obvious, compared to figure 9.}
\label{}
\end{center}
\end{figure}

\subsection{Spherical and plane parallel models}

An important conclusion of the previous section is that after finite shear the arcade expands and releases its energy. It is essential to have a seed poloidal arcade, where we can amplify its energy by almost a factor of ten by twisting before it erupts. This is in agreement to what \cite{1991ApJ...375L..61A} proved, that only finite energy can be stored in a 3-D force-free magnetic structure. The question of 2-D plane parallel structures and 3-D axially symmetric arcades is discussed in general by \cite{2002ApJ...574.1011U} where the physical deviations of the two configurations are stressed.  

A similar version of this problem has been solved by \cite{1988ApJ...324..574L, 1994A&A...288.1012A} where solar flares were described through a 2-D plane parallel model.  In order to have a direct comparison we construct a poloidal arcade of the same size in 2-D and we evaluate the expansion as a function of the horizontal shearing (Figure 11). Initially, both the 2-D and 3-D arcades, they coincide, however when the size of the 3-D arcade becomes comparable to that of the star their evolution changes drastically, the 2-D model continues on a linear increase of the size of the arcade with the horizontal shearing, whereas the expansion of the 3-D is much more rapid. Besides that, there is need for infinite amount of energy in order to force the plane parallel arcade to open and reach infinity as opposed to the finite energy demands of the 3-D arcade.

\begin{figure}
\begin{center} 
\epsfxsize=8.7cm 
\epsfbox{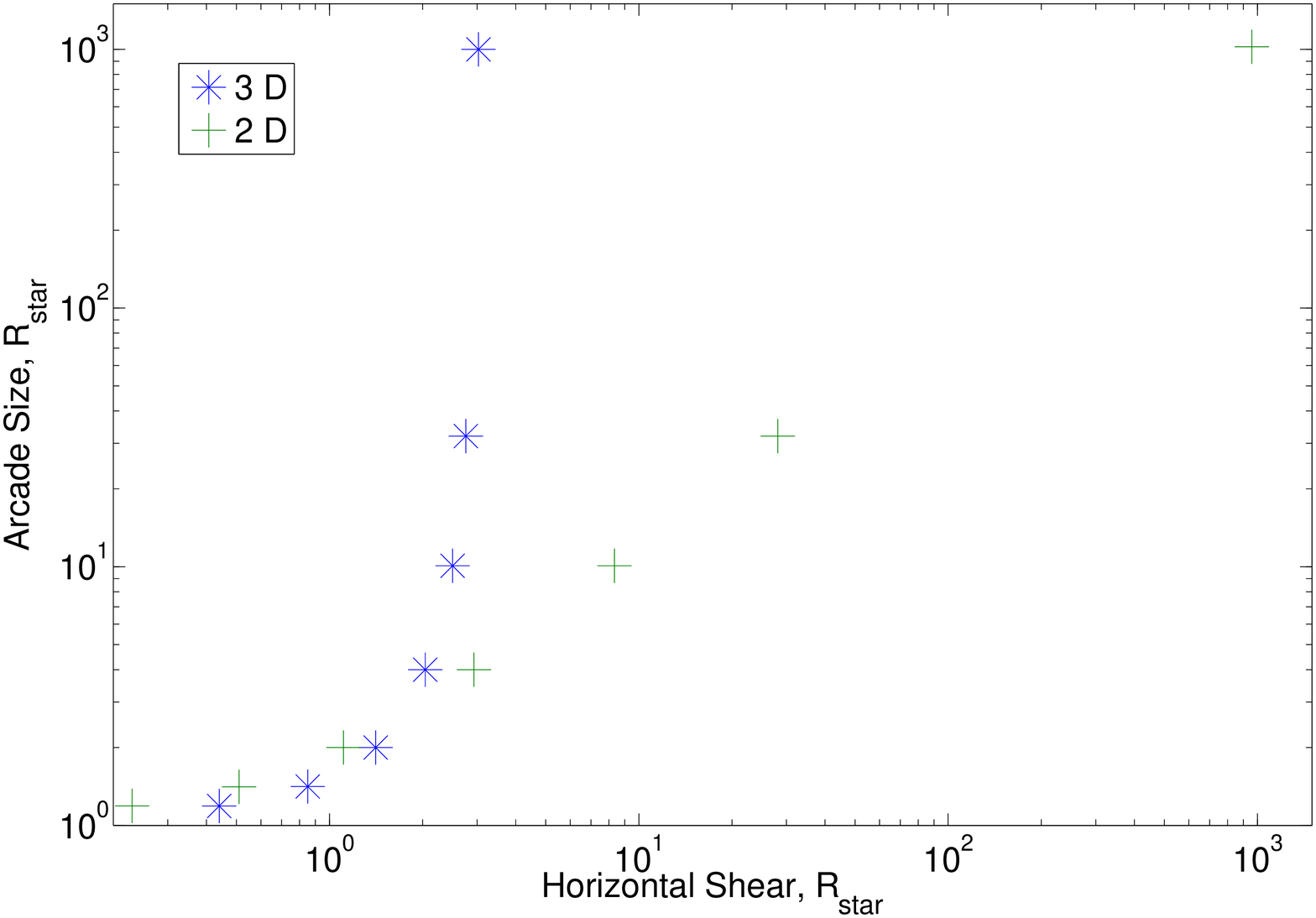}
\caption[]{The size of the arcade as a function of the horizontal shear in the 2 D and 3 D models. We start with two arcades of the same size, when the 3 D arcade becomes comparable to the size of the star their evolution changes dramatically.}
\label{}
\end{center}
\end{figure}

\section{The Pressure}

In this section we address the question of the pressure environment of the arcade. We estimate the dependence of the magnetic pressure on $r$ and how it varies with $l$. Afterwards we comment on the evolution in arbitrary environments, we do not deduce detailed solutions for the structure of the field inside the magnetic cavity, because an arbitrary environment does not allow analytical self-similar solutions and a detailed numerical model is beyond the scope of this paper; however based on physical arguments about pressure equilibrium at the edge of the magnetic cavity, we gain some insight about its shape. 

There is a physical reason that prevents the existence of self-similar solutions in the case of an arbitrary environment. If we include an arbitrary pressure we increase the external parameters and we are able to construct by the means of dimensional analysis a typical length scale; systems having a typical length scale do not allow self-similar solutions.

\subsection{Equilibrium with an external environment}

In order to confine this structure, it has to balance with an external pressure at its boundaries. The magnetic pressure $p_{m}$ at $\mu_{1}$ and $\mu_{2}$ is only due to the $B_{r}$ component of the magnetic field as the other two components vanish. 

\begin{eqnarray}
p_{m}(r)=\frac{1}{8 \pi}\Big(\frac{F \alpha^{l}}{2 \pi}f'(\mu_{1})r^{-l-2}\Big)^{2},
\end{eqnarray}

and by setting $p_{m,0}=\frac{1}{8 \pi}\Big(\frac{F \alpha^{l}}{2 \pi}f'(\mu_{1})\Big)^{2}$, the pressure is

\begin{eqnarray}
 p_{m}(r)=p_{m,0}r^{-2l-4}.
\end{eqnarray}

If the pressure of the gas surrounding the structure is obeying a power law $p_{g}\propto r^{-\zeta}$, then in order to have equilibrium the indexes must satisfy

\begin{eqnarray}
\zeta=2l+4
\end{eqnarray}

It is rather unphysical to expect a natural environment to obey this varying pressure law.  

It makes better sense to study this structure subject to a given pressure. But we have to sacrifice the advantage of self-similar analytical solutions and seek a numerical solution. It could be studied using methods similar to the ones developed by \cite{2006MNRAS...369.1167L}.

\subsection{Evolution under given pressure environment}

In section 4.1 we found that the magnetic pressure is proportional to $r^{-2l-4}$, and in our arcade models we set $l$ to decrease from 9 down to 0. Consider an environment where the pressure is proportional to $r^{-\zeta}$ and it is in equilibrium with the arcade at the surface of the star, thus $p_{g}(a)=p_{m}(a)$. For any distance $r_{1}$ that is $r_{1}>a$ the tendency of the arcade to expand or to contract in the angular sense depends on the sign of $Q=2l+4-\zeta$. If $Q$ is positive the gas pressure will be greater than the magnetic pressure, therefore, the arcade is more compressed and may indeed be pushed down and smothered. If $Q$ is negative the arcade expands in the angular direction, whereas for $Q=0$, the opening angle of the arcade will remain constant and the self-similar solution will satisfy the boundary conditions. 

In the cases studied, we assumed that $l$ varies from 0 to 9, therefore it is interesting to describe the behaviour of an arcade that evolves in an environment of certain $\zeta$ particularly when $\zeta>4$ and $\zeta=4$, the case of $\zeta<4$ is not discussed here because the system does not pass through a stage of self-similar solutions. The choice of $l=9$ initially constrains $\zeta<13$, otherwise it would be impossible to confine the arcade at any stage. 

Let $\zeta>4$, initially $l=9$ and $Q>0$ therefore the gas pressure collimates the arcade. However, as $l$ decreases, there is some value of $l$ for which $Q=0$, at this point there is no deviation from the self-similar solutions. Just after that $Q$ becomes negative, the opening angle of the arcade increases and the arcade opens to infinity. Therefore the system releases its energy faster than it does in the fully self-similar model.

For $\zeta=4$, $Q$ is always positive, unless $l=4$, therefore the arcade is always confined by its boundaries and its opening angle decreases as we move away from the surface of the star, the final part of its evolution is the split monopole, at this stage the arcade obeys the solution for $l=0$ and it consists of two streams of field lines, one outgoing and the other incoming that connect at infinity. 

The process can be summarised in the following stages. We start with a magnetic structure where the gas pressure is stronger than the corresponding pressure of a self-similar arcade. We inject energy to the system by shearing the foot points, and the arcade expands so that it reaches a stage where self-similar solutions are permitted, at this stage, the arcade is confined between two cones and is described by the solutions of the previous section. If we shear a bit more the foot points the arcade deviates from the self-similar solution and in addition to the radial expansion it opens in the angular direction. The amount of shearing and energy needed to achieve that, is defined by the dependence of the pressure to the distance and can be roughly approximated by the energy of the self-similar arcades appearing on table 1.

\section{Time Evolution}

The discussion of the problem did not include any time dependence. We are going to study this aspect of the problem in more detail. Using the equilibrium solutions we find the dependence of the differential rotation at the surface $\Phi$ on $l$. We are going to assume that $\Phi$ increases linearly with time, and then we are going to see how fast the arcades expand due to the displacement at the surface. At this point we are going to estimate the relation between twist and expansion for a given field line that corresponds to $P=1/2P_{max}$. Let the angular displacement of the foot points of a field line determined by $P$ be $\Phi_{P}$, and let the displacement imposed at the base increase linearly with time for this field line

\begin{eqnarray}
 \Phi_{P}=\Omega_{P}t
\end{eqnarray}

A motion at the origin $\delta \xi =a \sin \theta \delta \phi$ corresponds to an expansion $\delta r$. The expansion velocity cannot exceed the Alfv$\grave{e}$n velocity in the medium hosting the arcade. This is because the perturbation responsible for the expansion cannot travel faster than a magnetic wave. In order to make an evaluation of the velocity we assume that the field emerges from a spherical star of the size of the sun, its strength is $10^{2} Gauss$ and the particle density of the plasma inside the arcade is $10^{9}cm^{-3}$, corresponding to an Alfv$\grave{e}$n velocity of $6 \times 10^{8}cm/sec$ \citep{2002A&ARv..10..313P}, that is $2\%$ of the speed of light. The angular velocity of the differential rotation is $10^{-7}rad/sec$. Then we compute the expansion of the arcade corresponding to this shearing and we expect that both take place in the same time scale. Therefore we set a limit for the validity of the model of quasi static solutions. In table 3 we present the corresponding velocities and the relation between the shear and the expansion. We conclude that under the above assumption the magnetic configuration cannot be studied with this method when the power law index becomes less than $0.1$. At the final stages the expansion velocity would be extremely fast and the field would not have time to come to an equilibrium. The details of this critical point depend on the choice of the parameters of the problem. In any case no matter how slow the shearing is, the expansion at some point becomes infinitely fast, a choice of a very small $\Omega$ may only postpone this for later stages. Therefore the assumption of quasi static solutions is limited. This problem demonstrates that it is essential to seek for other solutions that include time evolution explicitly and create a fully dynamical model. This task seems far from the capacity of the current analytical model, however there are 3-D dynamical simulations \citep{2006A&A...460..909M} giving some insight.

\begin{table}
 \centering
 \begin{displaymath}
\begin{array}{|l|l|l|l|l|}
\hline
t(\times 10^{6}sec)& \Phi  & l   & r_{max} (R_{\sun}) & v_{exp} (cm/sec) \\ 
\hline
0                & 0        & 9  & 1.08            & 0.0 \\ 
1.7              & 0.52    & 4   & 1.19            & 4.4\times10^{4} \\ 
3.3              & 1.00    & 2   & 1.41            & 9.7\times10^{4} \\ 
5.5              & 1.66    & 1   & 2.00            & 1.8\times10^{5} \\ 
8.1              & 2.42    & 0.5 & 4.00            & 5.4\times10^{5}\\ 
10.9             & 3.28    & 0.2 & 32            & 6.8\times10^{6}\\ 
12.0             & 3.60    & 0.1 & 10^{3}           & 6.4\times10^{8}\\ 
12.6             & 3.77    & 0.05& 10^{6}           & - \\
\hline
 \end{array}
 \end{displaymath}

 \caption{The time evolution of the field line corresponding to $P=P_{1/2}$.}
 \label{Table 3}
\end{table}

\section{Conclusions}

In this study we looked in detail the possibility of constructing magnetic arcades. The Grad-Shafranov equation \citep{1958,1966} accepts self-similar solutions confined between two cones and of arcade topology. In order to avoid numerical computations we make certain assumptions about the distribution of the toroidal component of the magnetic field inside the structure. The starting point is a poloidal current-free magnetic field whereas the endpoint is the split monopole. At the split monopole stage a sheet current forms at the boundary between incoming and outgoing magnetic field, reconnection effects must be very important at this stage of evolution \citep{1996A&A...306..913A}. A similar case studied is that of the triple arcade, in this scenario we study a triple arcade focusing on the central arcade. The boundaries of the central arcade are allowed to move in latitude on the surface of the star. We conclude that there is no great difference between the two models, thus in future work, focusing on the single arcade is sufficient. 

The comparison of the 2-D and 3-D models is a very interesting topic. Two initially similar arcades, one in 2-D plane parallel model and the second in 3-D azimuthal symmetry, diverge as their size increases. The critical point for the divergence occurs when the 3-D arcade becomes comparable with the size of the star. 

We also address the question of the pressure. The analytical idealised solutions we obtain may only appear in certain pressure environments. If we look for a more realistic pressure environment we have to allow the boundaries of the arcade to vary with altitude and not to be confined between the two cones, however we have to sacrifice self-similarity and make a fully numerical model that is beyond the scope of this paper. However the use of our solutions provides essential insight about the evolution of such structures. 

Finally we studied the time evolution of this structure. We considered a field line corresponding to $P_{1/2}$ of the flux and we studied its evolution. Assuming that we make this shear under constant velocity at the base we expect that the expansion should take place in the same time scale. This time scale has to be shorter than the Alfv$\grave{e}$n crossing-time, otherwise the message of the change of the shear at the base will not be transmitted to the top of the arcade. Therefore making a rough estimate we can set an upper limit to the validity of this method, relating the shear velocity at the origin and other physical quantities of the magnetic structure. We concluded that there is a limit related to the physical properties of the system, the static solutions cannot be valid for ever, therefore a dynamical study of the problem is essential.

\section{Acknowledgements}

The author is grateful to Professor Donald Lynden-Bell for stimulating discussions and guidance and to the anonymous referee for insightful comments.

{}

\appendix{APPENDIX: THE TOTAL TWIST}

The twist of a field line emerging at and sinking to latitudes corresponding to $\mu_{1}$ and $\mu_{2}$ respectively is given by the integral of the equation (18). When $l \to 0$ the angular function $f(\mu)$ can be approximated by the relation (20), section 3.1. Let 

\begin{eqnarray}
w=\ln \cosh \Big(\frac{(l+1)c_{0}}{l(1-\mu_{max0}^{2})^{1/2}}(\mu-\mu_{max0})\Big),
\end{eqnarray}

\begin{eqnarray}
f^{1/l}=(1-\frac{l}{l+1}w)^{1/l} \approx \exp{(-w)},
\end{eqnarray}

therefore the integrand of equation (18) becomes

\begin{eqnarray}
\frac{c_{0}f^{1/l}}{l(1-\mu^{2})} \approx \frac{c_{0}}{l(1-\mu^{2})\cosh\Big(\frac{(l+1)c_{0}}{l(1-\mu_{max0}^{2})^{1/2}}(\mu-\mu_{max0})\Big)}.
\end{eqnarray}

The variation of the integrand of equation (18) due to $f^{1/l}(\mu)$ is much stronger than the variation due to $1-\mu^{2}$, since well away from $\mu_{max0}$ $f$ is zero. We put then $1-\mu^{2}=1-\mu^{2}_{max0}$ and we integrate. 

\begin{eqnarray}
\Phi=-\frac{c_{0}}{l(1-\mu^{2}_{max0})}\int \frac{d \mu}{\cosh\Big(\frac{(l+1)c_{0}}{l(1-\mu_{max0}^{2})^{1/2}}(\mu-\mu_{max0})\Big)}.
\end{eqnarray}

This integration can be done analytically and gives

\begin{eqnarray}
\Phi=-\frac{2}{\sqrt{1-\mu_{max0}}(l+1)}\{\arctan \exp[c_{1} (\mu_{2}-\mu_{max0})] \nonumber \\
- \arctan \exp[c_{1}(\mu_{1}- \mu_{max0})]\},
\end{eqnarray}

where $c_{1}=(l+1)c_{0}/(l\sqrt{1-\mu_{max0}})$. When $l \to 0$, $c_{1}$ becomes pretty big, therefore the first term in the bracket is $\pi/2$, whereas the second term is $0$. Thus, the total twist is:

\begin{eqnarray}
\Phi=-\frac{\pi}{\sqrt{1-\mu_{max0}^2}}.
\end{eqnarray}

\bsp

\label{lastpage}

\end{document}